\documentclass[
reprint,amsmath,amssymb,aps,pra,floatfix,]{revtex4-1}
\usepackage{graphicx}	
\usepackage{dcolumn}	
\usepackage{bm}		

\begin{document}

\title{Evaluation of the strength of electron-proton scattering data for determining the proton charge radius}

\author{M. Horbatsch and E.A. Hessels}
\email{hessels@yorku.ca}

\affiliation{%
 Department of Physics and Astronomy, York University, Toronto, Ontario M3J 1P3, Canada
}%

\date{\today}

\begin{abstract}
Precisely measured electron-proton elastic scattering 
cross sections [Phys. Rev. Lett. {\bf 105}, 242002 (2010)]
are reanalyzed to evaluate their strength for 
determining the rms charge radius 
($R_{\rm E}$) 
of the proton.
More than half of the cross sections at lowest $Q^2$ are 
fit using two single-parameter form-factor models, 
with the first based on a dipole parametrization, 
and the second on a linear fit to a conformal-mapping
variable. These low-$Q^2$ fits extrapolate the slope of 
the form factor to $Q^2$=0 and determine $R_{\rm E}$ values of approximately 
0.84 and 0.89~fm, 
respectively. 
Fits spanning all $Q^2$, 
in which the single constants 
are replaced with cubic splines
at larger $Q^2$,
lead to similar results for $R_{\rm E}$.
We conclude that the scattering data is consistent with
$R_{\rm E}$ ranging from at least 0.84 to 0.89~fm, and 
therefore cannot 
resolve the discrepancy between  
determinations of $R_{\rm E}$ made using muonic and electronic
hydrogen-atom spectroscopy.
\begin{description}
\item[PACS numbers]
\verb+\pacs{06.20.Jr,13.40.Gp,14.20.Dh,25.30.Bf}+
\end{description}
\end{abstract}

\maketitle

Recent measurements of the $n$=2 energy intervals of 
muonic hydrogen, when compared to precise 
QED theory for this exotic atom, lead to a determination 
\cite{pohl2010size,antognini2013proton}
of the rms charge radius of 
the proton 
($R_{\rm E}$) 
of 0.84087(39)~fm.
This value disagrees by 4.5 standard deviations 
with a value of 0.8758(77)~fm 
obtained from a similar comparison 
\cite{mohr2012codata} between QED theory 
and several precision measurements in
the ordinary hydrogen 
atom.
A third determination of $R_{\rm E}$ can be obtained
from precise measurements of the the cross sections for 
elastic scattering 
between electrons and protons.
The most precise electron-proton scattering experiment
is the recent measurement 
\cite{PRL.105.242001} of the MAMI collaboration,
and their analysis \cite{bernauer2014electric,*BernauerThesis} leads to 
$R_{\rm E}$= 0.879(8)~fm, 
which disagrees with the muonic hydrogen 
value by 4.6 standard deviations.

CODATA \cite{mohr2012codata} uses a combination of  
the scattering and hydrogen values to obtain 
$R_{\rm E}$, and its value differs from the 
muonic hydrogen value by 7 standard deviations.
This disagreement has now
widely been referred to as the 
proton size puzzle \cite{bernauer2014proton}. 
Many papers have discussed this puzzle,
including many that have proposed physics beyond the
standard model 
\footnote{See reviews of these discussions in Refs. 
\cite{carlson2015proton}
and
\cite{pohl2013muonicJ}.
}.

Because of the importance of the electron-proton scattering
data to this puzzle, the data has been extensively scrutinized
and discussed \cite{
PhysRevLett.107.119101,PhysRevLett.107.119102,
lorenz2012size,sick2012problems,
pohl2013muonicJ,adamuvsvcin2013advanced,
kraus2014polynomial,
bernauer2014electric,*BernauerThesis,lorenz2014reduction,
lorenz2015theoretical,carlson2015proton,
lee2015extraction,arrington2015evaluation,pacetti2015proton}. 
The present work reanalyzes the scattering
data and concludes that  it is consistent with a much larger range
of $R_{\rm E}$ values than obtained by others. 
This range 
makes it consistent with both the hydrogen and muonic hydrogen
determinations of $R_{\rm E}$, 
therefore removing one component of the 
proton radius puzzle.

Our analysis uses two separate simple 
one-parameter form-factor models.
Both of these simple models fit well to the low-$Q^2$ data,
but the two give discrepant values for
$R_{\rm E}$.
Since neither model can be ruled out, 
the uncertainty in $R_{\rm E}$ must,
at minimum, 
be expanded to encompass both values.
Generalizations of both models fit well to the entire MAMI data set,
and give similarly discrepant values for $R_{\rm E}$. 

The differential cross section for elastic scattering
of an electron of energy 
$E$ scattering by an angle 
$\theta$ 
from a stationary proton,
(after taking into account radiative corrections and
two-photon exchange)
can 
be written \cite{arrington2011review}
in terms of the squares of the electric and magnetic form factors 
($G_{\rm E}(Q^2)$ and
$G_{\rm M}(Q^2)$):
\begin{equation}
\sigma_{\rm red}=(1+\tau)\frac{{\rm d}\sigma}{{\rm d}\Omega} \Big/ \frac{{\rm d}\sigma_{\rm Mott}}{{\rm d}\Omega}=
G_{\rm E}^2+\frac{\tau G_{\rm M}^2}{ \epsilon }  ,
\label{eq:ratioToMott}
\end{equation}
where 
${\rm d}\sigma_{\rm Mott}/{\rm d}\Omega$ 
is the Mott differential cross section,  
$\epsilon$=$(1+2(1+\tau)\tan^2\frac{\theta}{2})^{-1}$, 
$\tau$=$-t/(4m_{\rm p}^2)$,
and
$t$=$-Q^2$=$(p_{\rm i}-p_{\rm f})^2$, 
with 
$p_{\rm i}$ and $p_{\rm f}$ 
being the initial and final four-momenta of the electron. 
Here, 
$m_{\rm p}$ is the proton mass,
and we use units with $\hbar=c=1$. 

In principle, the quantity of interest for this work,
\begin{equation}
R_{\rm E}=
\sqrt{3\frac{{\rm d}G_{\rm E}^2}{{\rm d}t}\Big{|}_{t=0}}
=\sqrt{3 \frac{\rm d \sigma_{\rm red}}{{\rm d}t}\Big{|}_{t=0}+\frac{3\mu_{\rm p}^2}{4 m_{\rm p}^2}},
\label{eq:REfromGE}
\end{equation}
could be determined using Eq.~(\ref{eq:ratioToMott}) 
from sufficiently-precise measurements of 
${\rm d}\sigma/{\rm d}\Omega$ for small values of 
$Q^2$.
In practice, 
for the existing set of measurements, 
an extrapolation to $Q^2$=0 is required, 
and, 
for this extrapolation, 
a functional form for $G_{\rm E}^2$ and $G_{\rm M}^2$ 
of Eq.~(\ref{eq:ratioToMott}) must be 
assumed.
 
The dipole form of the form factor has been
used for many decades 
\footnote{See, for example, L. N. Hand, D. G. Miller, and R. Wilson,
Rev. Mod. Phys. 35, 335 (1963)} and it approximates 
$G_{\rm E}$ and $G_{\rm M}/\mu_{\rm p}$ 
(where $\mu_{\rm p}$ is the magnetic moment 
of the proton in units of nuclear magnetons)
as:
\begin{equation}
G_{\rm E}^2=
\left(1+\frac{Q^2}{b_{\rm E}}\right)^{\!\!-4}
\!\!\!\!\!,\ \ \ \ 
\frac{G_{\rm M}^2}{\mu_{\rm p}^2}=
\left(1+\frac{Q^2}{b_{\rm M}}\right)^{\!\!-4}
\!\!\!\!\!.
\label{eq:dipole}
\end{equation}
A second approximation for the form factors is to use a 
Taylor expansion in $t$ about $t$=0. 
This Taylor expansion has a limited radius of convergence
due to a negative-$Q^2$ pole at $t$=4$m_{\pi}^2$ 
(where $m_{\pi}$ is the 
mass of the charged pion),
which results from the two-pion production threshold. 
A conformal mapping variable \cite{hill2010model}:
\begin{equation}
z=\frac{\sqrt{t_c- t}-\sqrt{t_c}}{\sqrt{t_c - t}+\sqrt{t_c}},
\label{eq:conformal}
\end{equation}
with $t_c$=4$m_{\pi}^2$ leads to a 
much larger radius of convergence.
Thus,
\begin{equation}
G_{\rm E}^2=1-c_{\rm E} z,
\ {\rm and}\ 
G_{\rm M}^2/\mu_{\rm p}^2=1-c_{\rm M} z
\label{eq:linearz}
\end{equation}
are good approximations to the form factors at low $Q^2$.

Other functional forms for 
$G_{\rm E}$ and $G_{\rm M}$ have been used,
to extrapolate to $Q^2$=0 to determine $R_{\rm E}$.
These other forms include 
polynomials in $t$ 
\cite{PRL.105.242001,bernauer2014electric,*BernauerThesis},
polynomials in $z$ \cite{lorenz2014reduction,lorenz2015theoretical,lee2015extraction},
inverse polynomials in $t$ \cite{PRL.105.242001,bernauer2014electric,*BernauerThesis},
dipole functions (Eq.~(\ref{eq:dipole})) times polynomials in $t$ \cite{bernauer2014electric,*BernauerThesis},
dipole functions plus polynomials in $t$ 
\cite{bernauer2014electric,*BernauerThesis},
cubic splines in $t$ 
\cite{PRL.105.242001,bernauer2014electric,*BernauerThesis},
dipole functions times cubic splines in $t$ 
\cite{bernauer2014electric,*BernauerThesis},
continued fractions in $t$ \cite{lorenz2012size},
and the Friedrich-Walcher parametrization 
(two dipole functions plus two symmetric
gaussian features) \cite{bernauer2014electric,*BernauerThesis}.
In this work, 
we restrict ourselves to the forms of Eqs.~(\ref{eq:dipole}) and (\ref{eq:linearz}) for 
low-$Q^2$ data,
and extensions of these forms for higher-$Q^2$ data.

\begin{figure*}
\includegraphics[width=7.0in]{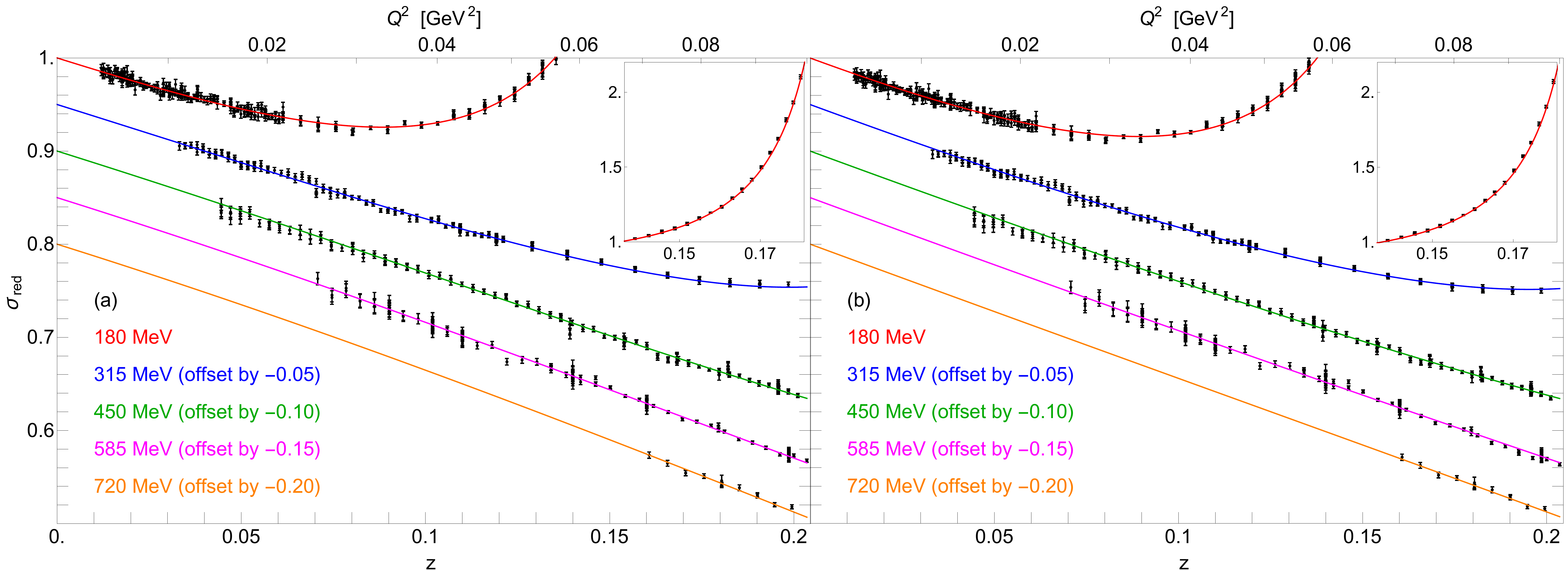}
\caption{\label{fig:lowQfits} (Color online) 
Fits to the low-$Q^2$ data of 
Ref. \cite{PRL.105.242001}
using single-parameter 
models for the form factors in Eq.~(\ref{eq:ratioToMott}). 
In (a), the dipole model of 
Eq.~(\ref{eq:dipole}) is used, and in (b) the linear model
of Eq.~(\ref{eq:linearz}) is used. 
Separate plots are required 
since the fits return different normalization 
constants, and therefore the cross sections take on 
slightly ($\sim$1\%) different values for the two plots. The $E$=180~MeV
data and fits are continued in the insets, 
and the other energies are offset for clarity of presentation.
}
\end{figure*} 

The highest-accuracy e-p scattering experiment 
\cite{PRL.105.242001} 
by the MAMI collaboration 
yields 1422 cross sections 
(with typical relative uncertainties of 0.35\%) 
spanning 
a range of 
180~MeV $\le E \le$855~MeV
and
16$^{\circ}\le\theta\le135.5^{\circ}$,
corresponding to 0.0038~GeV$^2 < Q^2 <$ 
1~GeV$^2$ 
and $0.06 < \epsilon < 1$.
The 1422 cross sections  
are divided into 34 data groups, 
with each data group having a separate 
normalization constant.
These normalization constants 
are known to an absolute accuracy of a few percent,
and are related to one another in such a 
way that there are only 31 independent constants
\cite{PRL.105.242001,bernauer2014electric,*BernauerThesis}.

The normalization constants add a further complication
to the $Q^2$=0 extrapolation needed to determine 
$R_{\rm E}$. 
The few-percent absolute accuracy of 
the measured cross sections is not
sufficient for performing a precise extrapolation, 
and thus
the 31 normalization constants
need to be 
floated 
when performing least-squares fits of the 
entire data set for this extrapolation.
We include these normalization constants in all of our fits.
Floating these constants adds considerable flexibility to the 
extrapolations. 
Although we do not impose the few-percent 
absolute uncertainty of the
normalization constants in our fits, 
all of our fits return constants near unity and well within this
few-percent uncertainty.

Other least-squares fits 
\cite{PRL.105.242001,bernauer2014electric,*BernauerThesis,
lorenz2014reduction,lorenz2015theoretical,
lee2015extraction,lorenz2012size} 
of this data use seven- to
twelve-parameter models for $G_{\rm E}$ and for $G_{\rm M}$,
and obtain least-squares fits with 
reduced $\chi^2$ values of as low as 1.14 for 
fitting the 1422 data points. 
The 1.14 value is much too large for the number of degrees of freedom in the fit,
but can easily be explained by either a 7\% underestimation of the uncertainties, 
or a systematic effect that is not fully accounted for. 
We only include fits that have a reduced $\chi^2$$<$1.14 in this work.

As indicated by Eq.~(\ref{eq:REfromGE}), 
the rms charge radius of the proton is a small-$Q^2$ concept.
Thus, 
if possible,
it should be determined from low-$Q^2$ data.
Therefore, 
we attempt to make a determination of $R_{\rm E}$ using 
fits to only the lower-$Q^2$ data. 
In addition to the fact that such fits use data nearer to $Q^2$=0,
the fits have the advantages that simpler, 
fewer-parameter models can be used 
for $G_{\rm E}^2$ and $G_{\rm M}^2$,
and that, 
since fewer of the data groups are used,
fewer normalization constants need to be included in the fits.

Fits using single-parameter models for the form factors
are shown in Fig.~\ref{fig:lowQfits}. 
These fits include data 
with $Q^2$$\le$$Q^2_{\rm max}$=0.1~GeV$^2$, from 
19 data groups, which require 17 normalization constants.
A data group is only included if there are more than 10 data points
in the group, and a total of 
761 of the 1422 cross sections (53\%) are used.
The value of $R_{\rm E}$ is directly obtainable from 
the slope of the curves in Fig.~\ref{fig:lowQfits}
at $Q^2$=0, and the fits provide the 
necessary extrapolation to $Q^2$=0. 

Fig.~\ref{fig:lowQfits}(a) shows a fit 
using Eq.~(\ref{eq:ratioToMott}) 
and the one-parameter dipole form factors of 
Eq.~(\ref{eq:dipole}).
The reduced $\chi^2$ for the fit is 1.11, 
and the fit returns 
$R_{\rm E}$=$(12/b_E)^{1/2}$=0.842(2)~fm
and
$R_{\rm M}$=$(12/b_M)^{1/2}$ =0.800(2)~fm.

A second fit to the same data uses the one-parameter linear model
(in $z$) of Eq.~(\ref{eq:linearz}).
This fit is shown in Fig.~\ref{fig:lowQfits}(b).
It also has a reduced $\chi^2$ of 1.11 and gives 
$R_{\rm E}$=$(\frac{3}{4} c_{\rm E}/t_c)^{1/2}$=0.888(1)~fm
and 
$R_{\rm M}$=$(\frac{3}{4} c_{\rm M}/t_c)^{1/2}$=0.874(2)~fm.

\begin{figure}[b]
\includegraphics[width=3.2in]{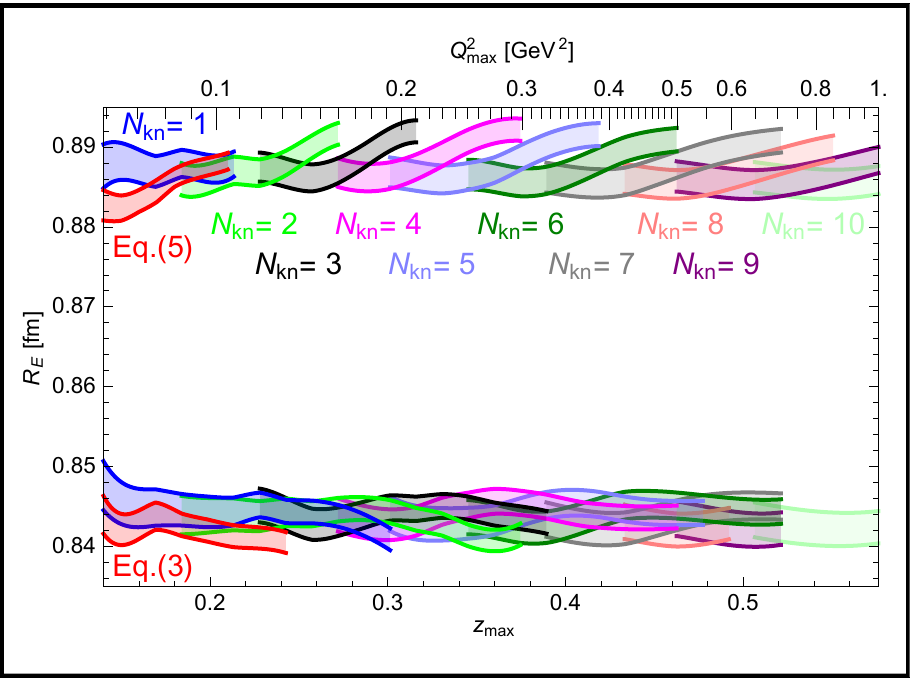}
\caption{\label{fig:REvsQsqr} (Color online) 
The range of $R_{\rm E}$ predicted from the 
dipole fits 
(lower bands) 
and conformal-mapping fits 
(upper bands) 
as a function
of the cutoff value $Q_{\rm max}^2$. 
The red bands correspond to single-parameter 
dipole and linear fits of 
Eqs.~(\ref{eq:dipole}) and (\ref{eq:linearz}),
and the other colors show extensions to these fits
for which the single parameters 
are replaced with cubic splines 
(with the number of nodes, $N_{\rm kn}$, as indicated)
at larger $Q^2$.
All fits shown have a reduced $\chi^2$$<$1.14. 
}
\end{figure}

Figure~\ref{fig:REvsQsqr} shows 
(red bands labeled Eq.~(\ref{eq:dipole}) and Eq.~(\ref{eq:linearz})
at the left of the figure) the error bands for 
$R_{\rm E}$ for the 
dipole and linear fits versus the cutoff $Q_{\rm max}^2$.
The figure includes the range of $Q_{\rm max}^2$ for which
 a reduced $\chi^2$$<$1.14 is obtained. 

\begin{figure}[b]
\includegraphics[width=3.0in]{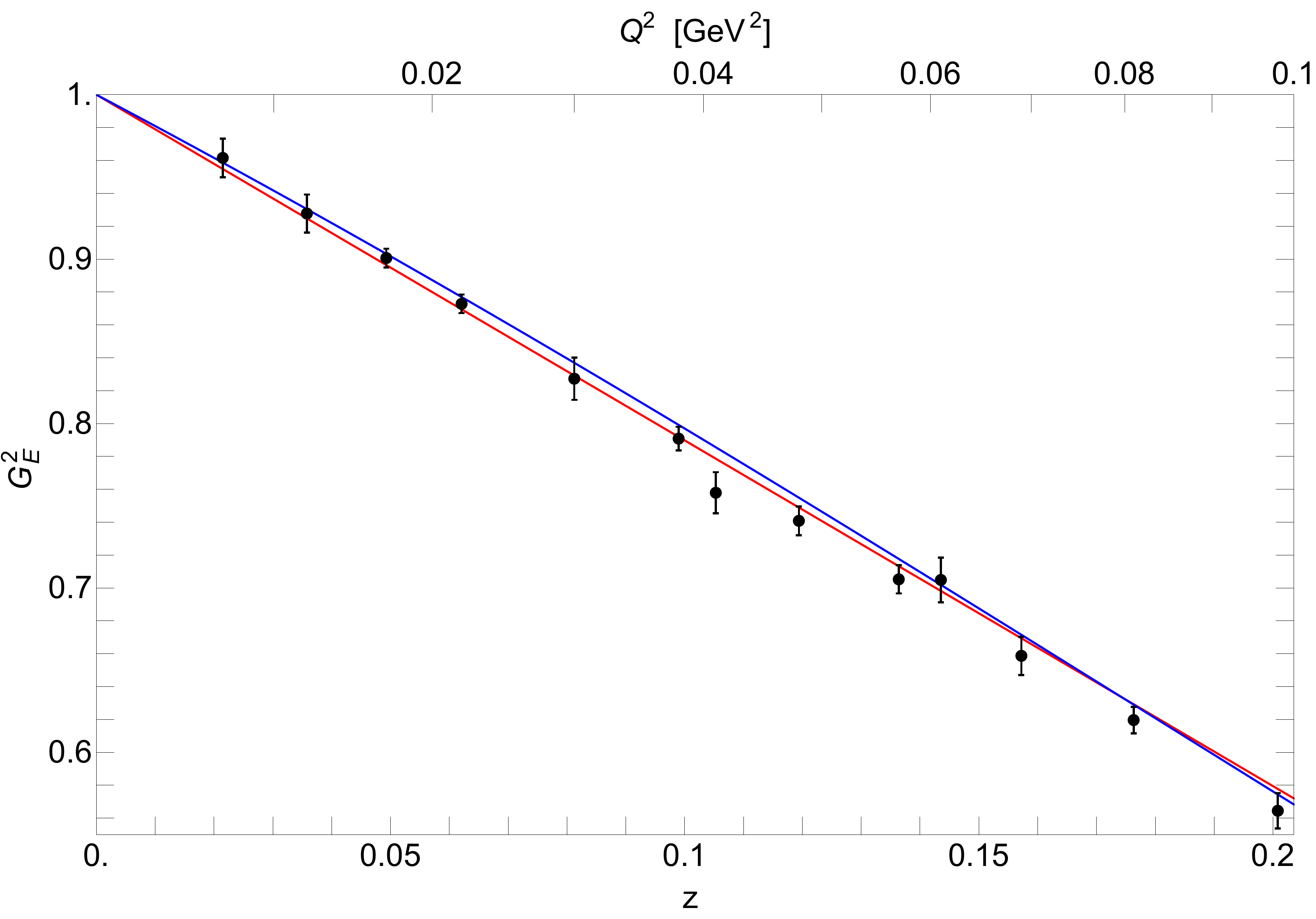}
\caption{\label{fig:lowQformFactors} (Color online) 
Electric form factor $G_{\rm E}$ determined 
from the single-parameter fits of Fig~\ref{fig:lowQfits}.
The blue curve is  from the dipole-model fit, 
and the red curve is from the 
linear-model fit. 
Also shown on the plots are the
low-$Q^2$ $G_{\rm E}$ data from other experiments
(from \cite{PRC.76.035205}), and these data 
are consistent with the $G_{\rm E}$ obtained
from either fit.
}
\end{figure}

The electric form factors predicted
from the two fits of Fig.~\ref{fig:lowQfits} 
are shown in Fig.~\ref{fig:lowQformFactors}.
Also plotted in the figure are other low-$Q^2$ measurements of 
these form factors (often referred to as the world data,
as summarized in Ref. \cite{PRC.76.035205}). 
It is clear from this figure (and from the
calculated $\chi^2$ for the comparison betwen the 
data and the two curves) that the form factor from 
either fit is also consistent with these other measurements.

One concern that could be raised about the 
single-parameter fits,
which are based only on data with $Q^2 < Q_{\rm max}^2$,
is that they may lead to inconsistencies 
for data with $Q^2$$>$$Q_{\rm max}^2$. 
Since the low-$Q^2$ fits of Fig.~\ref{fig:lowQfits} 
determine 17 of the 31  normalization constants,
and since data groups using these normalization 
constants include measured cross sections with 
$Q^2 > Q_{\rm max}^2$, 
the fits have a direct impact on data not included when fitting.
It could,
therefore,
be possible that data at the same value of 
$Q^2 > Q_{\rm max}^2$ from 
two of these data groups could be made inconsistent when these 
normalization constants are used.

To ensure that such inconsistencies do not take place, 
we extend the fits of the previous section to include all of the MAMI
data.
Such an extension also allows for a direct comparison of the quality of
our fits to the quality of the fits performed by others 
\cite{PRL.105.242001,bernauer2014electric,*BernauerThesis,
lorenz2014reduction,lorenz2015theoretical,
lee2015extraction,lorenz2012size}
who also include all of the MAMI data.
The extended fits include a range of $Q^2$ in which the functional 
form of $G_{\rm E}^2$ and $G_{\rm M}^2$ becomes more complicated,
and,
as with the fits performed by others,
more parameters are necessary to obtain a good fit.

The fit using Eq.~(\ref{eq:dipole}) can be generalized by 
allowing the constants $b_{\rm E}$ and $b_{\rm M}$ to become
functions of $z$. We do this by using cubic splines, 
with the $b$ values each being a constant for $z < z_{\rm 0}$=0.1,
and with $N_{\rm kn}$ equally spaced knots between 
$z_{\rm 0}$ and $z_{\rm max}$.
The values of $b$, 
and their first and second derivatives are continuous at 
$z_{\rm 0}$ and at the other knots. 
The number of knots needed to achieve
a good fit increases with increasing $z_{\rm max}$.

The results of such fits with $N_{\rm kn}$=1 to 10  
(2 to 11 parameters per form factor) 
are given in Fig.~\ref{fig:REvsQsqr}.
Again, only fits with 
a reduced $\chi^2$$<$1.14 are shown. 
The fits return values of $R_{\rm E}$ of approximately 
0.84~fm for all values of $Q^2_{\rm max}$,
similar to the single-parameter dipole fit. 
The fits at the right of the plot include all of the MAMI
data, and still return a value near 0.84~fm.

Equation (\ref{eq:linearz}) would predict 
negative values for 
$G_{\rm E}^2$ and $G_{\rm M}^2$ at larger $z$. 
This problem can be avoided by using a denominator
to cause a cutoff at higher $z$ values:
\begin{equation}
G_{\rm E}^2=\frac{1-c_{\rm E} z}{1-(c_{\rm E} z)^P}
\ {\rm and}\ 
\frac{G_{\rm M}^2}{\mu_{\rm p}^2}=\frac{1-c_{\rm M} z}{1-(c_{\rm M} z)^P}.
\label{eq:extendedLinear}
\end{equation}
We use $P$=4 for our fits, but any $P$ 
from 4 to 14 gives similar results. 
This function is very nearly linear up to $z$=0.2,
while avoiding negative values at larger $z$.
A fit of the data to these form factors is performed by 
allowing $c_{\rm E}$ and $c_{\rm M}$ to become functions of 
$z$ by using the same form of cubic splines as are used for
$b_{\rm E}$ and $b_{\rm M}$.

The results of such fits with $N_{\rm kn}$=1 to 10 
are also given in Fig.~\ref{fig:REvsQsqr}.
The fits return values of $R_{\rm E}$ of approximately 
0.89~fm, 
similar to the single-parameter linear fit.

Fig.~\ref{fig:REvsQsqr} clearly indicates that the 
two types of fits produce values of $R_{\rm E}$ that
disagree with 
each other. 
Since either type of fit gives an 
extrapolation to zero $Q^2$ that is equally valid, 
and since the quality of the fits are similar, 
either value of $R_{\rm E}$ is possible.
Therefore,
at best, 
the determined value of $R_{\rm E}$ can range
from 0.84 to 0.89~fm.
At worst, 
other valid extrapolations could lead to even 
a wider range of possible values,
leading to an even larger range for $R_{\rm E}$.

It is not the aim of this work to determine the 
rms magnetic radius of 
the proton, $R_{\rm M}$, but we note that the values from
our fits range from about 0.80 to 0.90~fm, 
and therefore
this work cannot determine $R_{\rm M}$ to any better than this
range. 
The consistency of the fits presented in this work can be 
checked by comparing the quantity 
$R_{\rm E}^2+R_{\rm M}^2$
to the prediction from hydrogenic spectroscopy.
The hydrogen hyperfine interval  
determines $R_{\rm E}^2+R_{\rm M}^2$ to be 1.35(12)~fm$^2$, and the
muonic hydrogen hyperfine interval determines
$R_{\rm E}^2+R_{\rm M}^2$ to be 1.49(18)~fm$^2$ \cite{PRD.90.053013}. 
The two determinations are consistent, and their
weighted average gives 
$R_{\rm E}^2+R_{\rm M}^2$ = 1.39(10)~fm$^2$.
The dipole fit of Fig.~\ref{fig:lowQfits}(a) gives 
$R_{\rm E}^2+R_{\rm M}^2$ = 1.349(4),
whereas the linear fit 
of Fig.~\ref{fig:lowQfits}(b) gives
1.553(4). The dipole fit is in excellent agreement
with the spectroscopy result, and the linear fit
shows only a mild 1.6 standard deviation discrepancy.
Similar comparisons using the extended cubic-spline 
fits lead to a similar level of agreement.

\begin{figure}[t]
\includegraphics[width=3.2in]{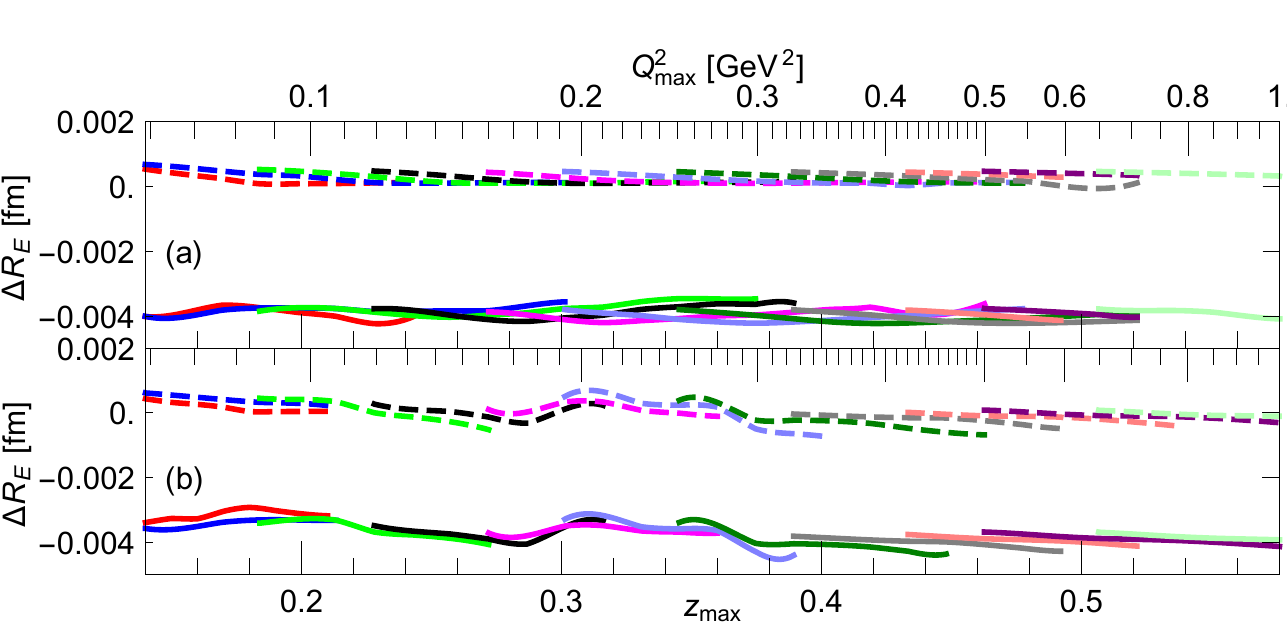}
\caption{\label{fig:tpe} (Color online) 
The sensitivity of the fit results of Fig.~\ref{fig:REvsQsqr} 
to two-photon exchange corrections 
is illustrated by redoing the analysis with 
poorer approximations for these corrections.
The change in $R_{\rm E}$ when
the Feshbach corrections (solid curves) and 
a low-$Q^2$ approximation 
\cite{borisyuk2007two} (dashed curves)
are applied shows that the radius extracted from both 
the dipole model (a) and conformal-mapping 
model (b) is not very sensitive to 
these corrections.
}
\end{figure}

The extent to which two-photon exchange (TPE) affects 
the extraction of $R_{\rm E}$ from the MAMI 
data has been debated in the literature
\cite{PhysRevLett.107.119101,
PhysRevLett.107.119102,
bernauer2014electric,*BernauerThesis,
pohl2013muonicJ,
lee2015extraction,
rachek2015measurement,
nikolenko2015proton,
tomalak2015two,
borisyuk2015two,
tomalak2015subtracted,
arrington2013coulomb,
adikaram2015towards}.
The cross sections given in Ref.~\cite{bernauer2014electric,*BernauerThesis} 
were corrected by the Coulomb corrections 
(Feshbach corrections
\cite{mckinley1948coulomb}) 
in place of the 
full TPE corrections.
In this work, these Coulomb corrections are removed 
and and replaced with TPE 
corrections calculated following the prescription
of \cite{borisyuk2012delta,borisyuk2012tpecalc}.
This replacement 
leads to correction factors of between 
0.997 and 1.003 for the data
of Fig.~\ref{fig:lowQfits}, and 
of between 0.978 and 1.003 
for the full MAMI set.
The correction factors agree 
with those shown in Fig.~5 of 
Ref.~\cite{lorenz2015theoretical} to
within the 0.03\% accuracy 
readable from their figure.
To test how sensitive our analysis is 
to TPE corrections, we repeat
our full analysis using the low-$Q^2$ TPE approximation
of Ref. \cite{borisyuk2007two} and the Feshbach
correction in place of the full TPE
correction. 
Fig.~\ref{fig:tpe} shows that 
using the Feshbach correction would
underestimate $R_{\rm E}$ by only 0.004~fm,
while the low-$Q^2$ approximation would change
$R_{\rm E}$ by less than 0.001~fm. 
We conclude that, 
although the best available TPE corrections should 
be used, the sensitivity to using poorer approximations
is small in our analysis.

In summary, we have reanalyzed electron-proton elastic
scattering data using simple fits to the lowest-$Q^2$ 
half of the data, and cubic-spline extensions of these fits
at higher $Q^2$. We find that the required
extrapolation to
$Q^2$=0  
can lead to values for the
rms charge radius $R_{\rm E}$ 
ranging from 0.84 to 0.89~fm.
This range does not resolve 
the discrepancy between determinations
of $R_{\rm E}$ from  
muonic hydrogen 
\cite{pohl2010size, antognini2013proton}
and ordinary hydrogen  
\cite{mohr2012codata}.

This work is supported by NSERC and CRC.

\bibliography{ProtonReferencesA}

\end{document}